\documentclass[reprint,amsmath,amssymb,aps,showkeys,nofootinbib]{revtex4-2}

\usepackage{graphicx} 
\usepackage{dcolumn} 
\usepackage{bm} 
\usepackage{multirow}
\usepackage{color}
\usepackage{xcolor}
\usepackage[colorlinks, urlcolor=blue, linkcolor=blue, citecolor=blue, plainpages=false]{hyperref}
\usepackage{bookmark}
\usepackage{subfig}
\usepackage{siunitx} 
\usepackage{mhchem}
\usepackage{diagbox}
\usepackage{tikz}
\usetikzlibrary{arrows.meta, positioning, angles, quotes}

\usepackage{comment}

\newcommand{\mr}[1]{\mathrm{#1}}

\begin{document}

\title{Extending the sensitivity of heavy sterile neutrino searches with solar neutrino experiments}%

\author{Yutao Zhu}
\author{Haoyang Fu}
\author{Wentai Luo}
\author{Shaomin Chen}
\author{Litao Yang}
\author{Zhicai Zhang}
\email[]{Correspondence: zhicaizhang@tsinghua.edu.cn}
\affiliation{%
 Department of Engineering Physics, Tsinghua University, Beijing 100084, China
}%
\affiliation{%
 Center for High Energy Physics, Tsinghua University, Beijing 100084, China
}%
\affiliation{Key Laboratory of Particle \& Radiation Imaging (Tsinghua University), Ministry of Education, China}

\date{\today}

\begin{abstract}
A sensitivity study of the search for heavy sterile neutrinos ($\nu_H$) in the MeV mass range using solar neutrino experiments is presented. $\nu_H$, with masses ranging from a few MeV up to around 15 MeV, can be produced in the Sun through $^8$B decay and subsequently decay into $\nu_e  e^+  e^-$. Its flux and lifetime strongly depend on the mixing parameter $|U_{eH}|^2$ and mass $m_{\nu_H}$. The $\nu_H$ signal can be detected via its decay products, either the $e^+e^-$ pair or $\nu_e$, depending on whether $\nu_H$ decays inside or outside the detector. Expected signal yields for both detection methods (detecting  $e^+e^-$ or $\nu_e$ signal) are presented across the full $|U_{eH}|^2$ and $m_{\nu_H}$ parameter space. These two methods are found to be complementary in different regions of the $|U_{eH}|^2$ and $m_{\nu_H}$ phase space. By combining both approaches, we anticipate observing at least a handful of signal events across most of the parameter space of $10^{-6} < |U_{eH}|^2 < 1$ and 2 MeV $< m_{\nu_H} < $ 15 MeV, assuming a 500-ton solar neutrino experiment operating for one year. Key variables, such as the energy spectra of $\nu_e$ or $e^+e^-$ and the $\nu_e$ solar angle, are also discussed to help distinguish signal from major backgrounds, such as solar neutrino events.
\end{abstract}

\maketitle

\section{\label{sec:introduction} Introduction }

The observation of neutrino flavor oscillations implies non-zero neutrino masses, 
marking one of the most significant discoveries in particle physics in the last few decades,
as it lies beyond the predictions of the Standard Model of particle physics (SM). 
Precisely determining neutrino masses and exploring new physics models to explain their non-zero values will remain a central focus of particle physics research in the coming decades.

Theoretically, only left-handed neutrinos were introduced in the SM at the time when the SM was established, as right-handed neutrinos were thought to be unnecessary and would be invisible (or "sterile," meaning no weak or electromagnetic interactions). 
Due to the absence of right-handed neutrinos, neutrinos are unable to acquire mass from Yukawa couplings as other fermions do; therefore, the SM predicts that all neutrinos are massless.
Experimentally, it is found that neutrino masses are not only non-zero but also exceptionally small: direct measurements from $\beta$ decay experiments~\cite{KATRIN2025} and indirect constraints from Cosmological surveys~\cite{DESI2024, Planck2018} both give an upper limit on the neutrino mass below $0.5~\mr{eV}$, which is more than six orders of magnitude smaller than the masses of other fermions in SM.

To explain the non-zero mass of neutrinos, new ingredients must be added to the SM, which can be either new particles, new symmetries, or other extensions. 
Among many neutrino mass models, one convenient way is to add right-handed neutrinos to the SM.
With right-handed neutrinos, neutrino masses can arise in various ways, such as Dirac masses as other fermions in the SM, or Majorana masses in the type-I seesaw mechanism~\cite{GellMann1979,Yanagida1979,Mohapatra1980}, where the former would require extremely small Yukawa couplings (less than $10^{-12}$) to explain the observed small neutrino masses, while the latter allows for larger couplings by introducing heavy sterile neutrinos that effectively suppress active neutrino masses. 

Although sterile neutrinos with extremely large masses (much larger than weak scale) are preferable to bring neutrino Yukawa couplings to a similar level as for other fermions in SM, heavy sterile neutrinos with mass in the range from $\mr{eV}$ to $\mr{TeV}$ are also appealing in many aspects. 
Besides being accessible to a wide range of experiments, they can also address other puzzles in particle physics, such as the nature of dark matter~\cite{ASAKA2005} and the mechanism of leptogenesis~\cite{Akhmedov1998}.

In the type-I seesaw model, the mixing between active neutrino $\nu_{\alpha}$ and sterile neutrino $\nu_H$, $|U_{\alpha H}|^2$, is approximately the ratio of the active and sterile neutrino mass, $m_{\nu_{\alpha}}/m_{\nu_{H}}$. 
Since the mass of active neutrinos is observed to be very small (less than $\sim 0.5~\mr{eV}$),
the mixing parameter would be extremely small for heavy sterile neutrinos, which means very tiny production rates of sterile neutrinos with usual neutrino sources (solar, reactor, accelerator, etc.), especially in the $\mr{MeV}$ and above range.
Some variants of the seesaw model, however, can make large mixing parameters possible (for example, reference~\cite{Xiaogang2009}).
As a result, searches for heavy sterile neutrinos in the $\mr{eV}$ to $\mr{TeV}$ mass range have been performed with a wide range of experiments, from collider experiments to beta decay measurements. 
Current results have excluded a wide range of $|U_{\alpha H}|^2$ and $m_{\nu_{H}}$ parameter space~\cite{Bolton2020}, but there is still an ample uncovered phase space up to the seesaw limit ($|U_{\alpha H}|^2\sim m_{\nu_{\alpha}}/m_{\nu_{H}}$).
The purpose of this paper is to discuss methods to further extend the sensitivity in the $\mr{MeV}$ mass range with solar neutrino detectors (currently in operation or under construction).

\section{\label{sec:productionAndDecay} Production and decay of heavy sterile neutrino}

Heavy sterile neutrinos ($\nu_H$) with masses of a few MeV can be produced from solar neutrinos ($\nu_e$) through active-sterile neutrino mixing, for example, from $^8 \mr{B}$ solar neutrino as a primary source:
\begin{equation}
    ^8\mr{B} \rightarrow ~ ^8\mr{Be}^* + e^+ + \nu_e\rightarrow ~ ^8\mr{Be}^* + e^+ + \nu_H
\end{equation}

\begin{figure}[!t]
\centering
\includegraphics[width=1.0\columnwidth]{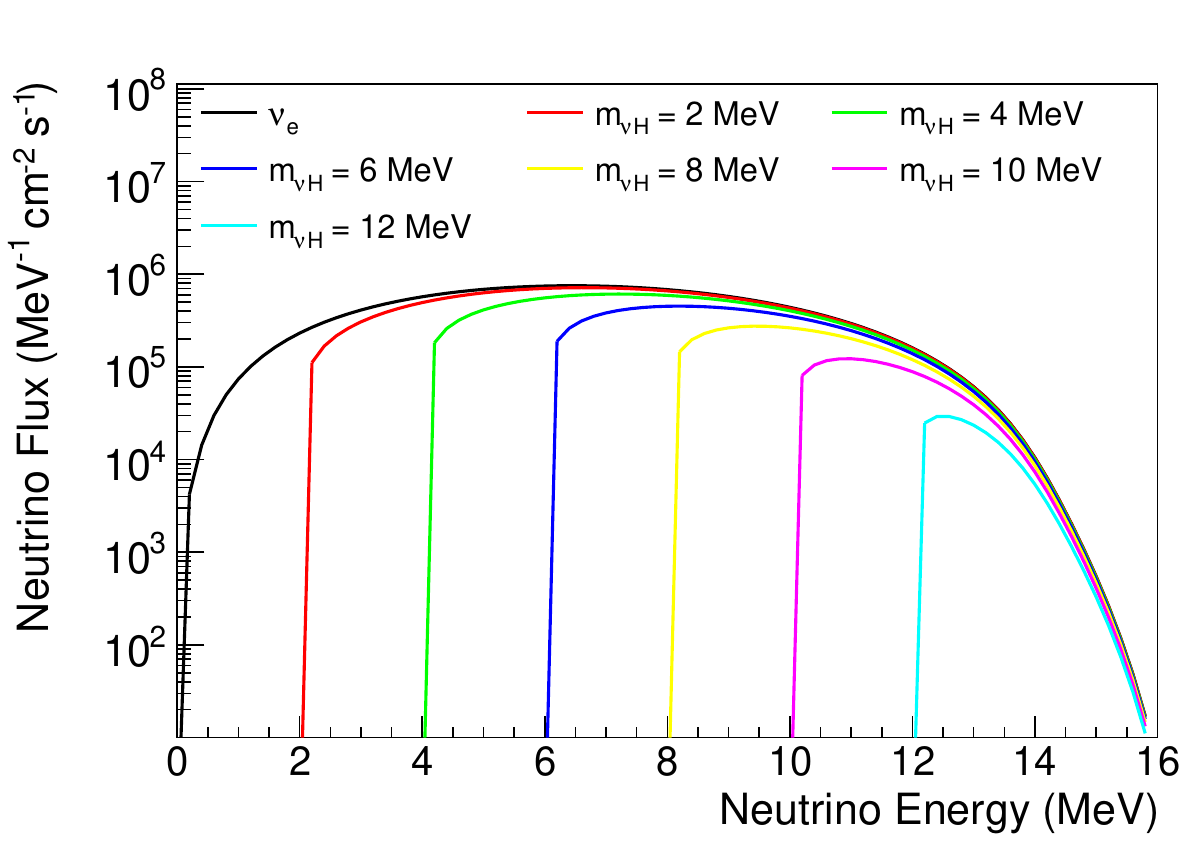}
\caption{{Energy spectra of $\nu_H$ with different masses emitted from $^8$B decay in the Sun. The spectra are based on the $^8 \mr{B}$ left-handed solar neutrino spectrum ($\nu_e$ in the plot, taken from~\cite{BLA+StandardNeutrinoSpectrum1996}) and are suppressed by the mixing parameter $|U_{eH}|^2$ and the phase-space factor according to Eq.~\ref{eq:nuHSpectrum}, where $|U_{eH}|^2 = 1.0$ in the spectra shown in this plot.}}
\label{fig:RHNSpectra} 
\end{figure}

The flux of $\nu_H$ from such decay chain is proportional to the $^8 \mr{B}$ solar neutrino flux and the mixing parameter $|U_{\alpha H}|^2$, scaled by a mass-dependent phase-space factor:

\begin{equation}
\label{eq:nuHSpectrum}
\Phi_{\nu_H}(E) = |U_{eH}|^2 \sqrt{1 - \bigg(\frac{m_{\nu_H}}{E}\bigg)^2}\Phi_{^8 \mr{B}}(E),
\end{equation}
where $m_{\nu_H}$ and $E$ are the mass and energy of the heavy sterile neutrino, respectively. 
Using Eq.~\ref{eq:nuHSpectrum} and $^8 \mr{B}$ solar neutrino spectrum from~\cite{BLA+StandardNeutrinoSpectrum1996}, we can plot the energy spectra of $\nu_H$ from $^8 \mr{B}$ decay for different $\nu_H$ masses, as shown in Fig.\ref{fig:RHNSpectra}.

The goal of this work is to explore possible methods to extend the discovery potential of such heavy sterile neutrinos with mixing parameter $|U_{eH}|^2$ much smaller than 1, getting as close as possible to the seesaw limit (for MeV mass, the seesaw limit is around $|U_{eH}|^2\sim10^{-7}$ or even smaller). 
With such a small mixing parameter, the deficit of $^8 \mr{B}$ solar neutrino due to mixing to sterile neutrino will be practically invisible to a solar neutrino detector of a few hundred tons (which has a daily $^8 \mr{B}$ solar neutrino event rate of about one). 
As a result, the rest of this paper will discuss the direct search for heavy sterile neutrinos from $^8 \mr{B}$ decay.

\begin{figure*}[!t]
\centering
\includegraphics[width=0.3\textwidth]{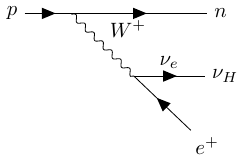}
\includegraphics[width=0.3\textwidth]{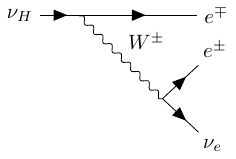}
\includegraphics[width=0.3\textwidth]{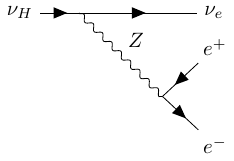}
\caption{Feynman diagrams of $\nu_H$ production from solar $^8 \mr{B}$ decay and its decay $\nu_H \rightarrow e^+e^-\nu_e$ via $W^{\pm}$ or Z boson exchange.}
\label{fig:feynman_nuHProductionDecay}
\end{figure*}

For $\nu_H$ with a mass heavier than two times the electron mass, they can decay into an $e^+e^-$ pair plus a left-handed neutrino $\nu_e$.
The Feynman diagram for the production of a heavy sterile neutrino from $^8 \mr{B}$ solar neutrino and its subsequent decay can be seen in Fig.\ref{fig:feynman_nuHProductionDecay}.

\begin{figure}[!h]
\centering
\includegraphics[width=1.0\columnwidth]{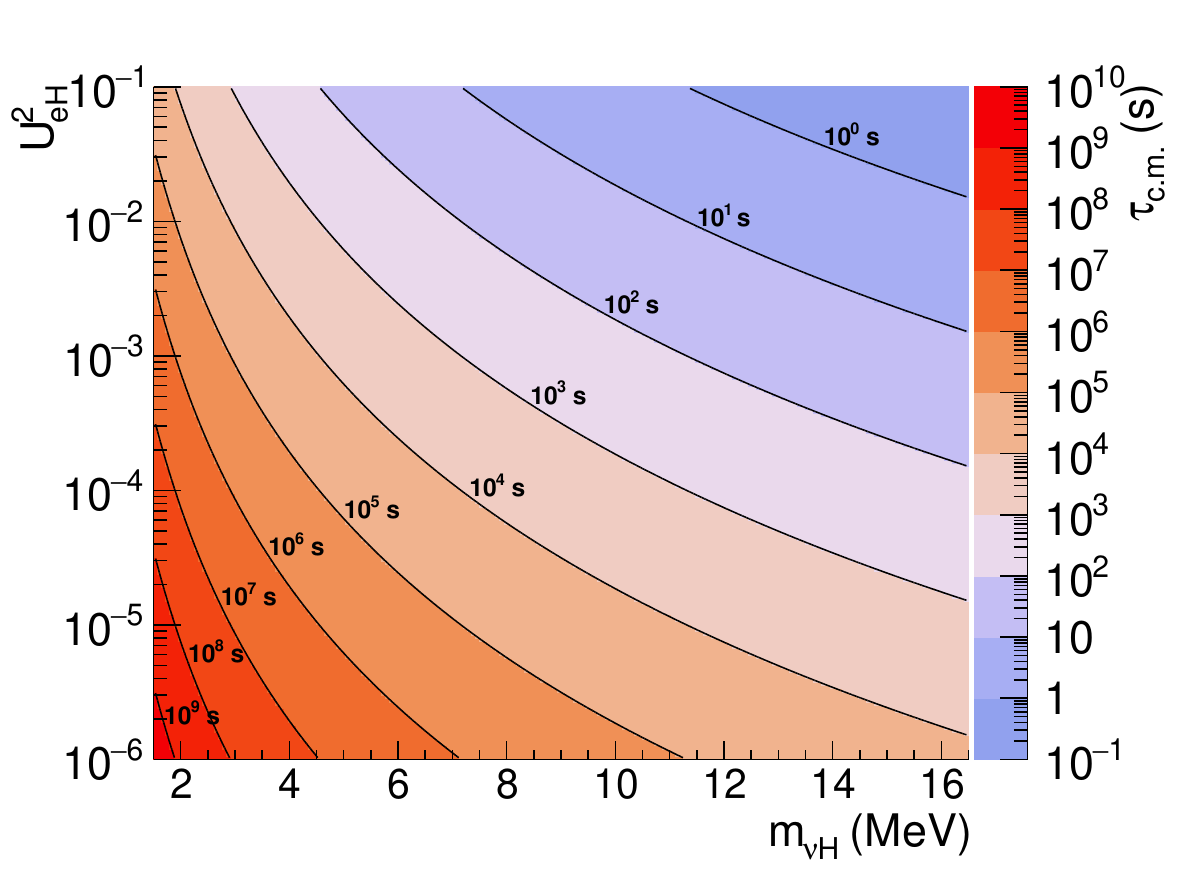}
\caption{{Proper lifetime ($\tau_{c.m.}=1/\Gamma_{\mr{total}}$) of $\nu_H$ as a function of mass $m_{\nu_H}$ and mixing parameter $|U_{eH}|^2$.}}
\label{fig:CTauRHN}
\end{figure}

In addition to $\nu_H \rightarrow e^+e^-\nu_e$ decay, $\nu_H$ can also decay into three neutrinos ($\nu_e$) via Z boson exchange, as well as radiatively decay into $\nu_e+\gamma$ through a $W$ and lepton loop. 
The width of $\nu_H \rightarrow \nu_e+\gamma$ decay is negligible compared to $e^+e^-\nu_e$ or three-neutrino decay.
Calculating the matrix element for $\nu_H$ decaying to three active neutrinos is straightforward, and after phase space integration, we obtain the width in the following form:
\begin{equation}
    \label{eq:Gamma_nuHto3nu}
    \Gamma_{3\nu_e} = \frac{G_F^2}{192\pi^3}m_{\nu_H}^5|U_{eH}|^2
\end{equation}
The width of $\nu_H \rightarrow e^+e^-\nu_e$ is similar to that of $\nu_H \rightarrow 3\nu_e$, with minor differences due to the mixed vector and vector-axial terms and the non-negligible electron mass. The exact value of $\Gamma_{e^+e^-\nu_e}$ can be found in~\cite{GSHowFindNeutral2007}. For $\nu_H$ with sufficiently large mass ($m_{\nu_H} \gg m_e$), the ratio of $\Gamma_{e^+e^-\nu_e}$ to $\Gamma_{3\nu_e}$ is approximately:
\begin{equation}
    \label{eq:ratio_Gammas}
    \frac{\Gamma_{e^+e^-\nu_e}}{\Gamma_{3\nu_e}} \approx \frac14 (1-4\sin^2\theta_w + 8\sin^{4}\theta_w) \approx 0.6
\end{equation}
From the width calculations above, we can determine the proper lifetime of $\nu_H$ decay as a function of its mass and the mixing parameter $|U_{eH}|^2$, as shown in Fig.\ref{fig:CTauRHN}. 
It can be seen that $\nu_H$ with small mass and small mixing parameter can have an extremely long lifetime, which is evident from Eq.~\ref{eq:Gamma_nuHto3nu}.
For the lower left corner (small mass and small mixing parameter), only a small fraction of $\nu_H$ decays before reaching our detector on Earth;
while for the upper right corner (large mass and large mixing parameter), most $\nu_H$ decays within Earth's orbit.

\begin{figure}[!htbp]
\centering
\includegraphics[width=1.0\columnwidth]{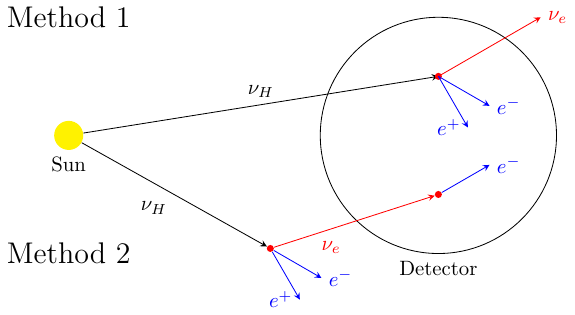}
\caption{Two detection Methods for heavy sterile neutrinos: (1) $e^+e^-$ pair detection from $\nu_H$ decays occurring inside the detector, and (2) $\nu_e$ detection from $\nu_H$ decays occurring outside the detector.}
\label{fig:diagram_scenarios}
\end{figure}

Since $\nu_H$ itself doesn't produce detectable signals in most traditional neutrino detectors (except those designed to detect extremely weak signals like coherent elastic neutrino-nucleus scattering, as reported in~\cite{XENON8B}), direct searches for $\nu_H$ typically rely on detecting its decay products. These could be either the $e^+e^-$ pair or the $\nu_e$ shown in Fig.\ref{fig:feynman_nuHProductionDecay}, as illustrated in Fig.\ref{fig:diagram_scenarios}:
\begin{itemize}
    \item \textbf{Method 1}: If $\nu_H$ decays inside the detector, we can search for the $e^+e^-$ signal event within the detector.
    \item \textbf{Method 2}: If $\nu_H$ decays before reaching the detector, the detectable signal would be a $\nu_e$ event in our detector.
\end{itemize}
The remainder of this paper explores both methods and discusses their complementary sensitivity across the full phase space of $m_{\nu_H}$ and $|U_{eH}|^2$.

\section{\label{sec:Method1} Method 1: search for $\nu_H$ by $e^+e^-$ signal}

To estimate the sensitivity of $\nu_H$ searches using the $e^+e^-$ signal from its decay, we calculate and present the event rates for both signal and major background events, along with the distribution of a key discriminating variable—the total energy deposition spectrum—across the entire $m_{\nu_H}$-$|U_{eH}|^2$ parameter space under investigation.

The signal event rate can be estimated by calculating the $\nu_H$ decay rate within the detector. For $\nu_H$ with production rate $R_0$ at the Sun (source), lifetime $\tau$, Sun-to-Earth time of flight $t$, time of flight inside the detector $\mr{d}t$, and detector solid angle $\mr{d}\Omega$ relative to the Sun, the $\nu_H$ decay rate inside the detector is given by:
\begin{equation}
\label{eq:RofTau}
    R = R_0 \frac{\mr{d}\Omega}{4\pi}\frac{\mr{d}t}{\tau}\exp\left(\frac{-t}{\tau}\right)
\end{equation}

To analyze the rate variation across different regions of the $m_{\nu_H}$-$|U_{eH}|^2$ parameter space, we can treat $\tau$ as the primary variable of change since $t$ and $\mr{d}t$ remain relatively constant for different $m_{\nu_H}$ and $|U_{eH}|^2$ values. By taking the derivative $\mr{d}R/\mr{d}\tau$ of Eq.~\ref{eq:RofTau}, we find that $R$ increases with $\tau$ until $\tau=t$ (approximately 500 seconds in most cases), after which $R$ decreases as $\tau$ continues to increase.

Combining this observation with the lifetime distribution shown in Fig.\ref{fig:CTauRHN} and the source rate ($R_0$) from Eq.~\ref{eq:nuHSpectrum}, we can conclude:
\begin{enumerate}
    \item The ratio $R/R_0$ (total signal efficiency) peaks near the $\tau=500~\mr{s}$ contour in Fig.\ref{fig:CTauRHN} and reaches minima at both corners (lower left and upper right).
    \item According to Eq.~\ref{eq:nuHSpectrum}, $R_0$ is maximized in the upper left corner of Fig.\ref{fig:CTauRHN} and minimized in the lower right corner.
    \item Consequently, the signal event rate $R$ contour lines in the $|U_{eH}|^2$ vs. $m_{\nu_H}$ plane resemble a crest aligned with the $\tau=500~\mr{s}$ contour (approximately diagonal in Fig.\ref{fig:CTauRHN}), with the maximum at the upper left corner and minimum at the lower right corner.
\end{enumerate}

\begin{figure}[!t]
\centering
\includegraphics[width=1.0\columnwidth]{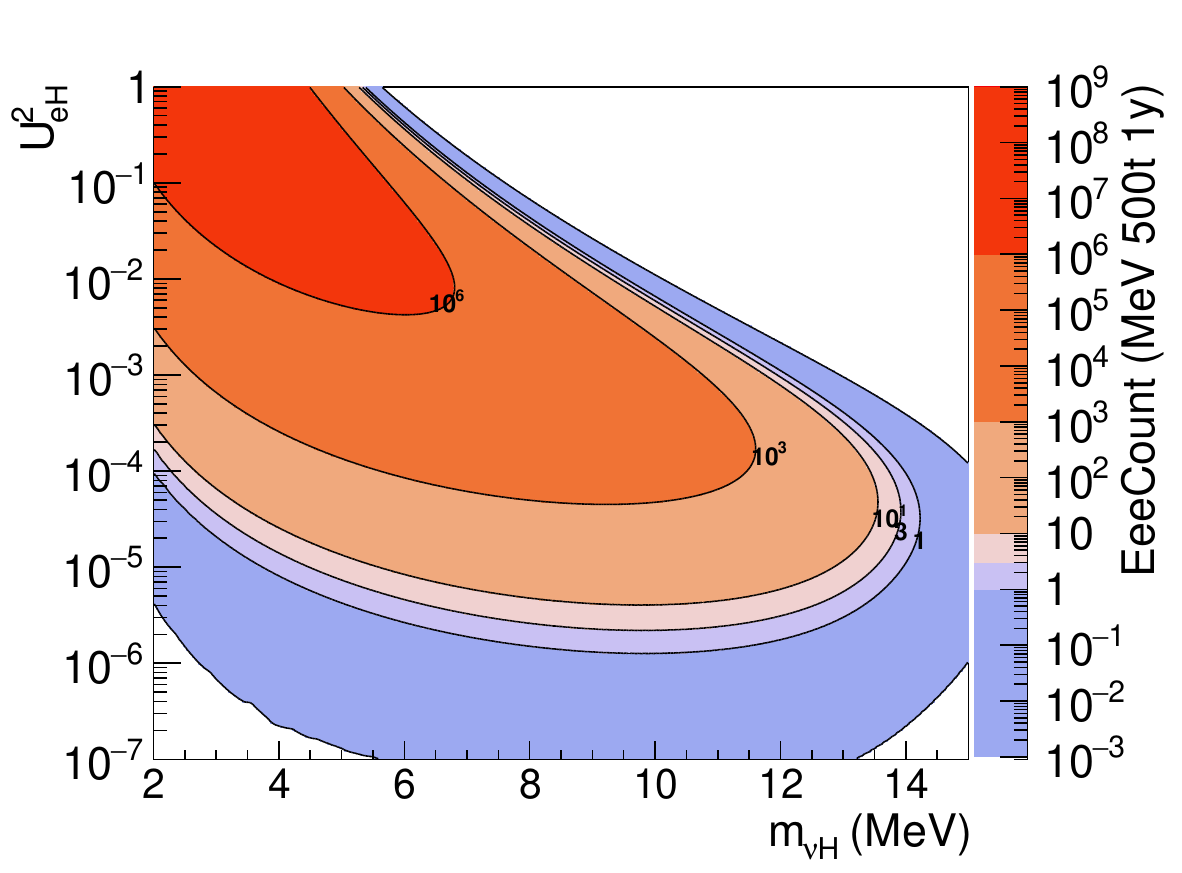}
\includegraphics[width=1.0\columnwidth]{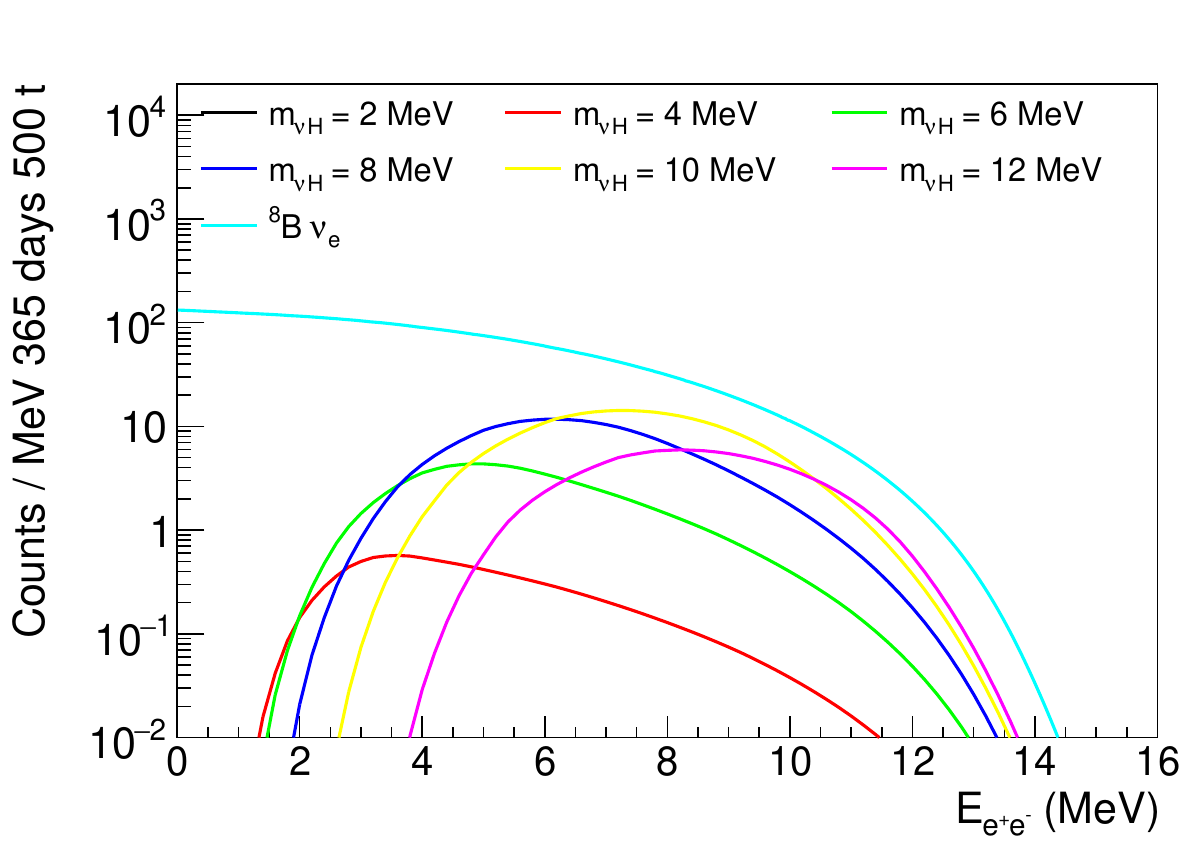}
\caption{Top: The count rate (per year) of $e^{+}e^{-}$ signal from $\nu_H$ decay inside a 500-ton detector on Earth, as a function of $m_{\nu_H}$ and $|U_{eH}|^2$. {Bottom: The energy spectrum of $e^{+}e^{-}$ signal from $\nu_H$ decay, along with the energy spectrum of the major background (scattered electron in the detector from solar $^8\mr{B}$ neutrino). Different values of $\nu_H$ mass are shown, with $|U_{eH}|^2 = 10^{-5}$.}}
\label{fig:eeSignalRateAndSpectrum} 
\end{figure}

The exact values of $R$ in Eq.~\ref{eq:RofTau} can be obtained by substituting $R_0$ with the total flux from Eq.~\ref{eq:nuHSpectrum} and calculating the time of flight ($t$ or $\mr{d}t$) using:
\begin{equation}
\label{eq:time_of_flight}
    t = \frac{m_{\nu_H}}{E_{\nu_H}}\frac{D}{\beta c}
\end{equation}
where $D$ represents the travel distance (Sun-Earth distance for $t$ or detector size for $\mr{d}t$). For a 500-ton spherical detector on Earth, we calculated the expected $e^+e^-$ signal event rates from $\nu_H$ decay across various $m_{\nu_H}$ and $|U_{eH}|^2$ values, as summarized in table~\ref{tab:eeSignalRate}.
These results are visualized in the top panel of Fig.\ref{fig:eeSignalRateAndSpectrum}, showing the expected crest-like contour pattern in the signal rate distribution.

\begin{table*}[!t]
\centering
\begin{tabular}{c|cccccc}
\hline
\diagbox{$|U_{eH}|^2$}{$M_{\nu_H}$ {/ MeV}} & 2 & 4 & 6 & 8 & 10 & 12\\
\hline
\num{e-1} & \num{9.7e5} & \num{3.8e7} & \num{2.4e5} & 0 & 0 & 0\\
\num{e-2} & \num{1.0e4} & \num{1.2e6} & \num{2.4e6} & \num{3.6e4} & 0 & 0\\
\num{e-3} & \num{1.0e2} & \num{1.5e4} & \num{1.1e5} & \num{1.2e5} & \num{9.2e3} & 2.2\\
\num{e-4} & 1.0 & \num{1.5e2} & \num{1.4e3} & \num{3.7e3} & \num{3.2e3} & \num{4.5e2}\\
\num{e-5} & 0 & 1.5 & 14.1 & 43.5 & 58.0 & 25.8\\
\num{e-6} & 0 & 0 & 0.1 & 0.4 & 0.6 & 0.3 \\
\hline
\end{tabular}
\caption{Expected event rate (per year) of $e^+e^-$ pairs from $\nu_H\rightarrow\nu_e e^+ e^-$ decays occurring within the detector, shown for different mixing parameters $|U_{eH}|^2$ and $\nu_H$ masses, assuming a 500-ton {neutrino} detector.}
\label{tab:eeSignalRate}
\end{table*}

The primary background for $e^+e^-$ signals from $\nu_H$ decays in solar neutrino detectors comes from electrons produced through elastic scattering of solar neutrinos with detector materials. 
The key distinction between $\nu_H$ signals and solar neutrino backgrounds lies in their final states: $\nu_H$ decays produce $e^+e^-$ pairs, while background events generate single electrons.

For Cherenkov detectors, $e^+e^-$ pairs can potentially be distinguished from single electrons when their opening angle is sufficiently large.
Scintillation detectors, however, must rely on differences in total deposited energy spectra for signal-background separation.
Advanced slow scintillation detectors with directional sensitivity~\cite{Luo_2023} can leverage both energy spectra and opening angle information to optimize signal-to-background ratios.

We now calculate the expected energy spectrum of $e^+e^-$ pairs from $\nu_H$ decays and compare it with solar neutrino background spectra.
The differential width for $\nu_H\rightarrow e^+e^-\nu_e$ decay can be found in~\cite{ShrGeneralTheoryWeak1981}, which takes the form:
\begin{equation}
\label{eq:diffGammaETheta}
\frac{\mr{d}\Gamma}{\mr{d}E\mr{d}\cos\theta} = \Gamma_{\mr{total}}(f_1 + \zeta |\vec{P}|f_S \cos\theta),
\end{equation}
where $E$ and $\theta$ represent the energy and emission angle of $\nu_e$ in the $\nu_H$ rest frame; $\Gamma_{\mr{total}}$ is the total decay width; $f_1$ and $f_S$ are functions of $m_{\nu_H}$ and $E$; $|\vec{P}|\approx \beta$ denotes the $\nu_H$ polarization; and $\zeta = +1~(-1)$ for $\nu_H$ ($\bar{\nu}_H$).

Using the $\nu_e$ energy spectrum from Eq.~\ref{eq:diffGammaETheta} in the $\nu_H$ rest frame, we compute the total $e^+e^-$ energy spectrum in the laboratory frame.
The $e^+e^-$ spectral shape varies with $m_{\nu_H}$, as shown in the bottom panel of Fig.~\ref{fig:eeSignalRateAndSpectrum}.
This plot compares $e^+e^-$ spectra from $\nu_H$ decays (normalized to rates in a 500-ton {neutrino} detector) with electron spectra from $^8\mr{B}$ solar neutrino elastic scattering.
Distinct spectral peaks emerge at the tail of the background spectrum for most $\nu_H$ masses, suggesting that precise energy measurements in scintillation detectors should enable excellent signal-background discrimination through spectral fitting.

Combining the information from the top and bottom plots of Fig.~\ref{fig:eeSignalRateAndSpectrum}, 
we expect to observe at least a handful of $\nu_H\to e^+e^-\nu_e$ events by detection of $e^+e^-$ inside the detector for most regions where $|U_{eH}|^2>10^{-6}$ and $2~\mr{MeV}<m_{\nu_H} < 15~\mr{MeV}$.
Using the $e^+e^-$ total energy spectrum, the Borexino experiment has excluded some phase space in the above region of $|U_{eH}|^2$ and $m_{\nu_H}$ using 446 days of Borexino data~\cite{BorexinoRHN2013}.
Future solar neutrino observatories with larger central volumes and direction measurement capabilities—and therefore better separation of $e^+e^-$ from single electron events—such as the Jinping Neutrino Experiment (JNE)~\cite{JNELOI2017,Luo_2023}, 
could further extend the sensitivity to $\nu_H$ in these regions.

The uncovered regions in the phase space are the lower-left and upper-right corners of $|U_{eH}|^2$ and $m_{\nu_H}$, as shown in Fig.~\ref{fig:eeSignalRateAndSpectrum},
where $\nu_H$ is either too short-lived or too long-lived,
leading to insufficient $\nu_H$ decays inside the detector.
To cover those regions, one must search for $\nu_H$ that decays outside the detector,
for which the only method is to detect the $\nu_e$ from $\nu_H$ decay,
which will be discussed in the next section.

\begin{figure}[!t]
    \centering
    \includegraphics[width=1.0\columnwidth]{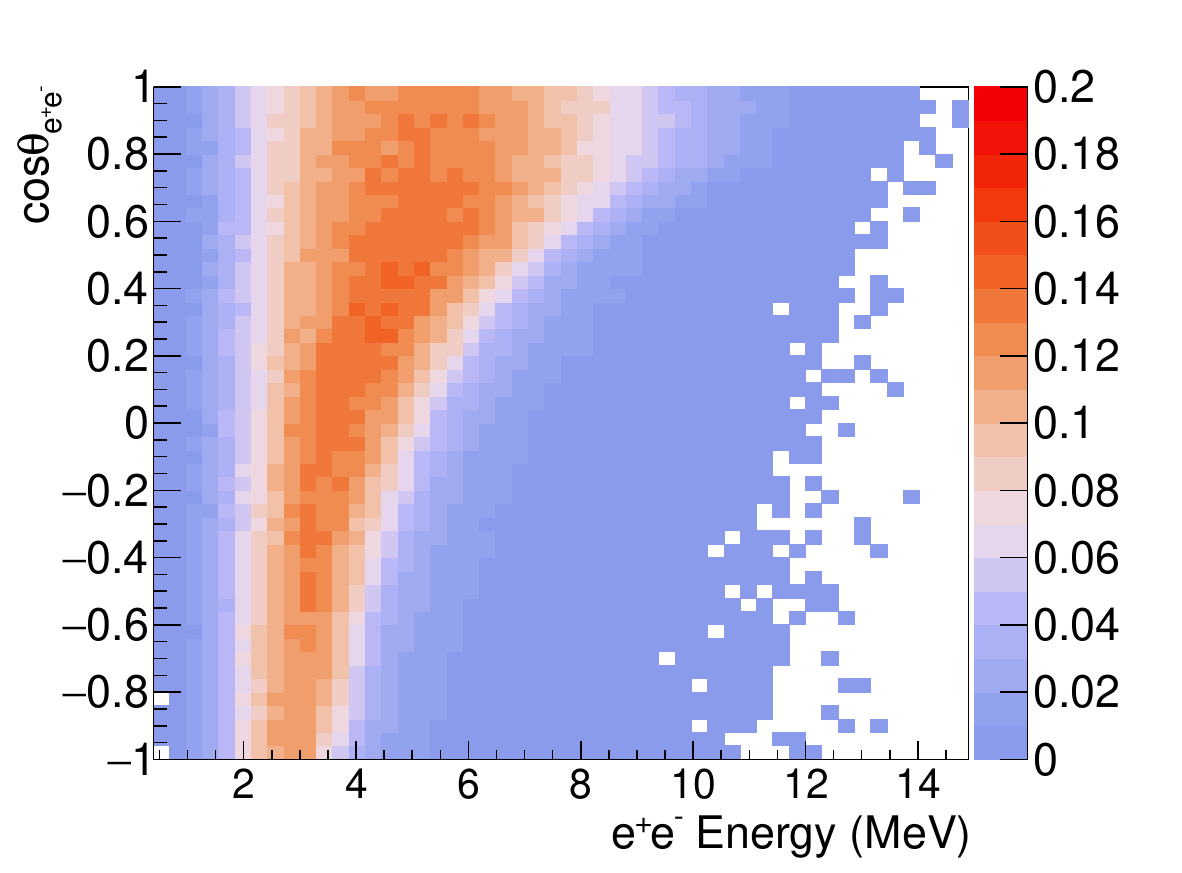}
    \includegraphics[width=1.0\columnwidth]{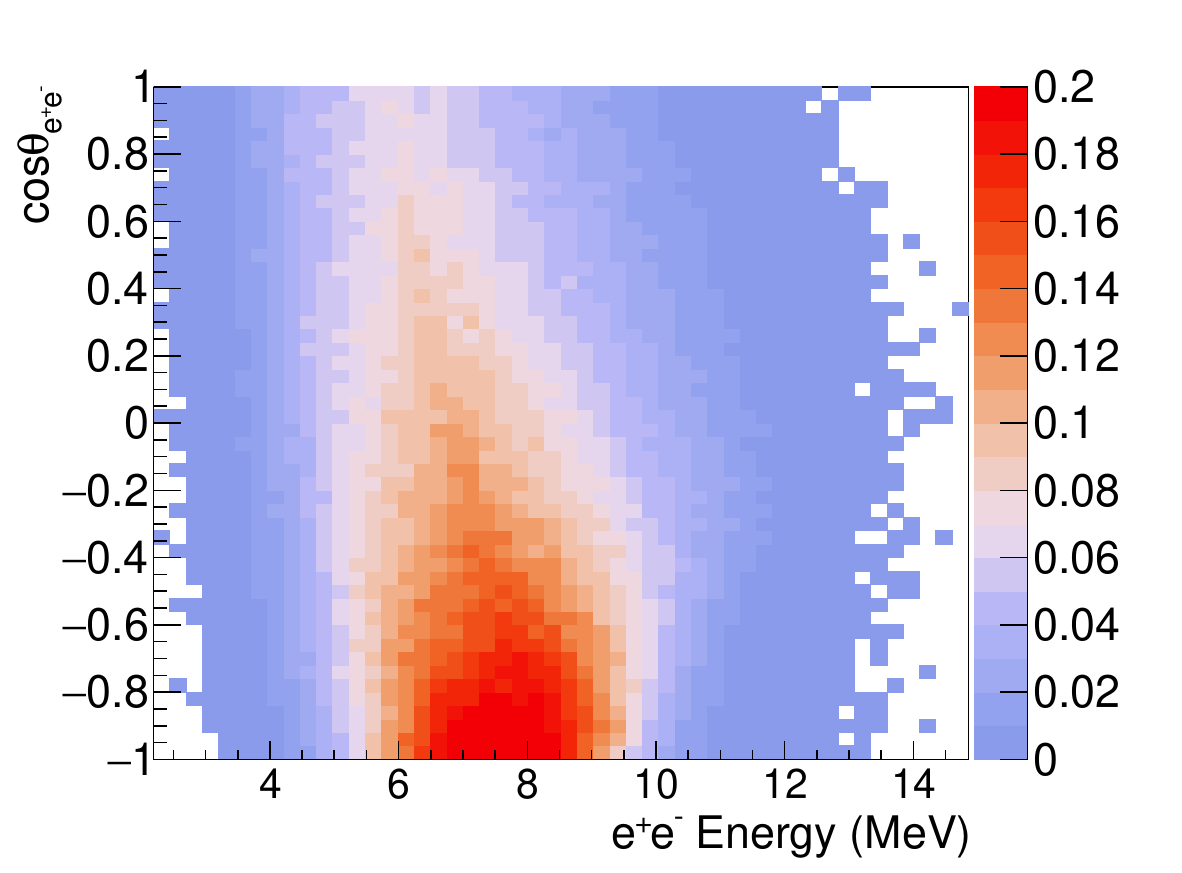}    
    \caption{Distribution of opening angle and energy of $e^+e^-$ pairs from $\nu_H$ decay with $|U_{eH}|^2=\num{e-05}$. Top: $m_{\nu_H}=\qty{4}{MeV}$; Bottom: $m_{\nu_H}=\qty{10}{MeV}$}
    \label{fig:s1_ee_cosine_dist}
\end{figure}

Furthermore, to obtain the energy and angular distributions of the decay products, we employ MadGraph aMC@NLO~\cite{alwall2014automated} to generate decay events in the rest frame of the heavy neutrino ($\nu_H \to \nu_e e^+ e^-$). These events are subsequently boosted to the laboratory frame by sampling from the predicted $\nu_H$ energy spectrum (Fig.~\ref{fig:RHNSpectra}). As shown in Fig.\ref{fig:s1_ee_cosine_dist}, for the majority of events, the opening angle of the $e^+e^-$ pair is sufficiently large to allow these pairs to be distinguished from background events containing single electrons. Moreover, as the mass of the right-handed neutrino increases, the proportion of events with a larger opening angle also increases.

\section{\label{sec:Method2} Method 2: search for $\nu_H$ by $\nu_e$ signal}

For regions of $|U_{eH}|^2$ and $m_{\nu_H}$ where the $\nu_H$ is too short-lived, the majority of $\nu_H$ will decay before they reach the Earth orbit.
Among the decay products, the only particle that can reach our detector on Earth is the active neutrino $\nu_e$.
The $\nu_e$ from $\nu_H$ decay is similar to solar neutrinos, making it challenging to distinguish this signal from the solar neutrino background. 
No such search has been performed with existing solar neutrino detectors worldwide.

This paper explores two variables to differentiate $\nu_e$ from $\nu_H$ decay and solar neutrino background: the $\nu_e$ energy and direction.
The signal $\nu_e$ is from $\nu_H$ decay, and $\nu_H$ itself is from a solar neutrino by active-sterile mixing; therefore, the energy of the signal $\nu_e$ should be significantly lower than the original solar neutrino energy.
As for the $\nu_e$ direction, the solar neutrino we detect on Earth always comes from the Sun, but the $\nu_H$ can decay anywhere in space and generate a $\nu_e$ that travels to the Earth. 
Consequently, the angle between the $\nu_e$ direction and the Sun-Earth line ($\theta_{\mr{Sun}}$) should be different for the signal and background.

\begin{figure}[!t]
\centering
    \includegraphics[width=0.85\columnwidth]{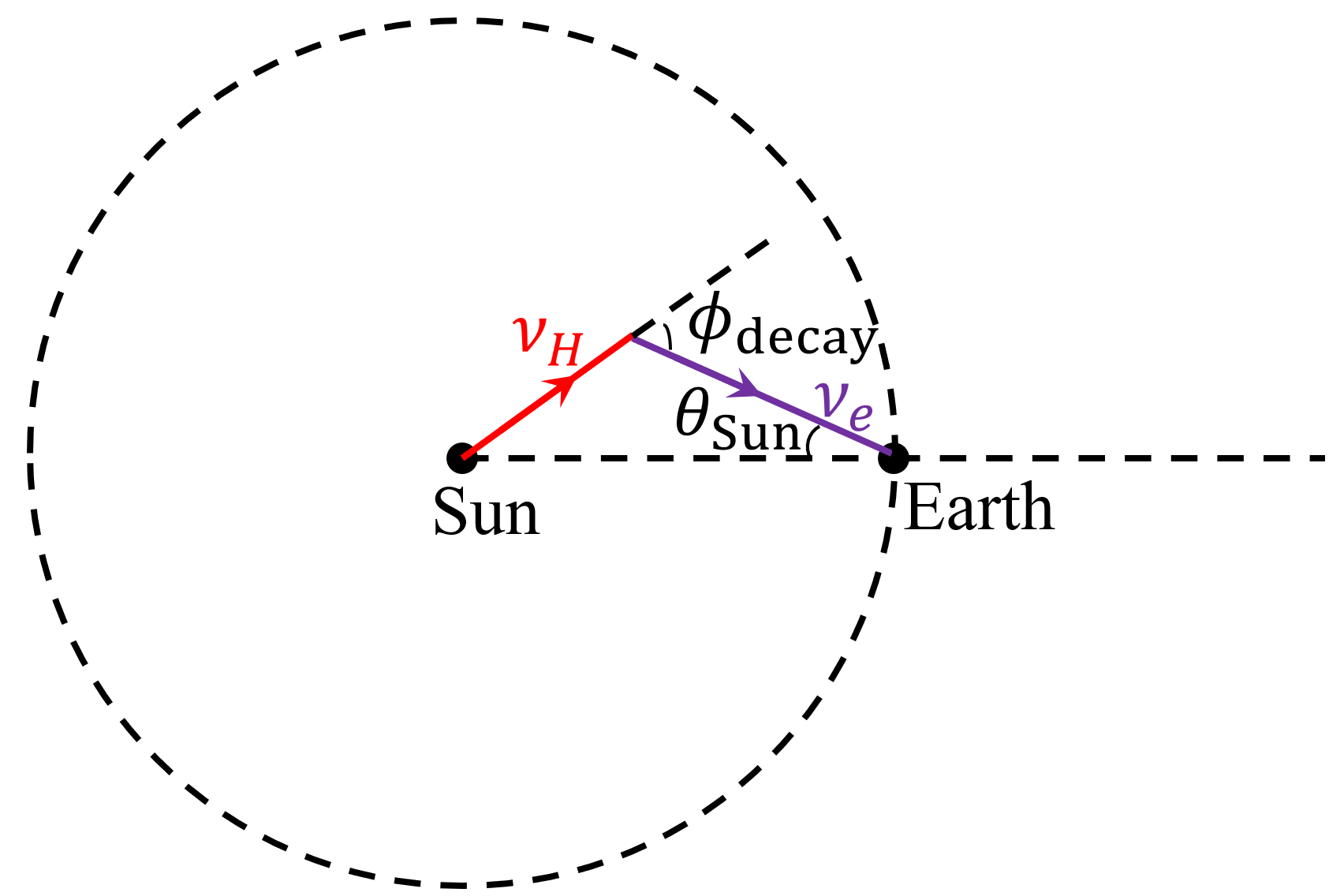}
    \includegraphics[width=0.85\columnwidth]{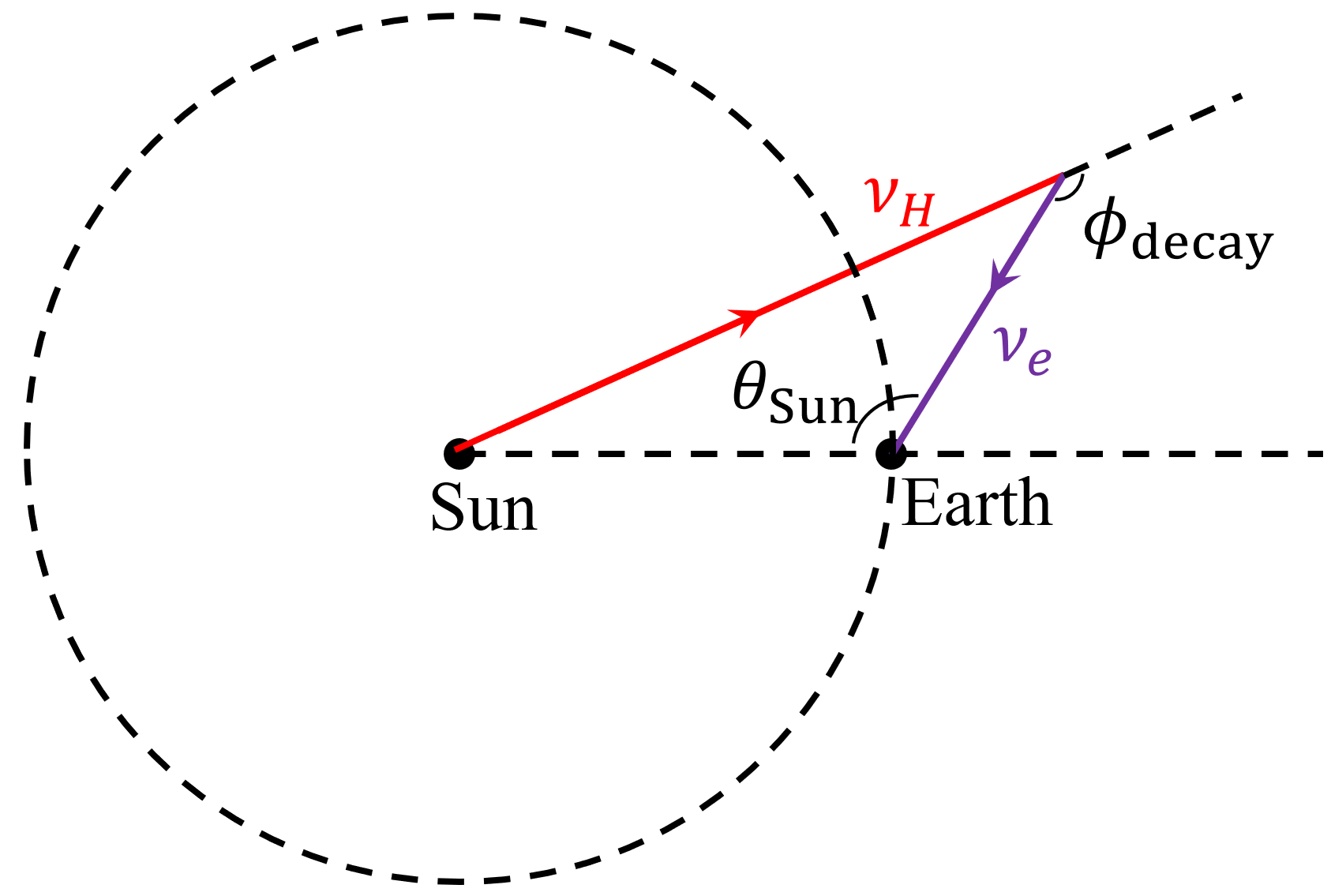}
\caption{Definition of angles $\phi_{decay}$ (emission angle) and $\theta_{Sun}$ (solar angle) for $\nu_H$ decay in flight and then the decay product $\nu_e$ reaches the detector on earth. Different Methods of $\nu_H$ decays are shown: $\nu_H$ decay inside Earth orbit (top plot), and $\nu_H$ decay outside Earth orbit (bottom plot).}
\label{fig:decay_in_flight_sketch}
\end{figure}

The calculation of the energy and angular distribution of $\nu_e$ from $\nu_H$ that reaches the detector on Earth 
begins with the energy and angular distribution of all emitted $\nu_e$ from $\nu_H$, as shown in Eq.~\ref{eq:diffGammaETheta}.
From Eq.~\ref{eq:diffGammaETheta}, we just need to integrate over all $\nu_H$ decay points in space and all $\nu_e$ emission angles to collect all $\nu_e$ that can reach our detector on Earth.
Fig.~\ref{fig:decay_in_flight_sketch} shows two typical situations of the decay of $\nu_H$ with $\nu_e$ entering the detector on Earth,
where the $\nu_H$ can either decay inside or outside the Earth's orbit.
Two angles are defined  and illustrated in Fig.\ref{fig:decay_in_flight_sketch}:
\begin{enumerate}
    \item $\nu_e$ emission angle $\phi_{\mr{decay}}$: angle between $\nu_H$ and $\nu_e$ direction. This angle is the same as the angle $\theta$ in Eq.~\ref{eq:diffGammaETheta} (except that $\theta$ in Eq.~\ref{eq:diffGammaETheta} is in $\nu_H$ rest frame, and $\phi_{\mr{decay}}$ is in lab frame). 
    \item $\nu_e$ solar angle $\theta_{\mr{Sun}}$: angle between $\nu_e$ direction and the Sun-Earth line.
\end{enumerate}

The 3-dimensional integration over all space can be simplified to a 1-dimensional integration when calculating the energy and $\theta_{\mr{Sun}}$ distributions of all $\nu_e$ that reach our detector on Earth.
The simplification is based on the fact that all detectors placed on any point on the spherical surface $S$ with the Sun as the center and the Sun-Earth distance as the radius should have the same energy and $\theta_{\mr{Sun}}$ distributions of $\nu_e$.
Therefore, we only need to consider decay vertices along a straight line from the Sun to infinity, 
integrating over decays where $\nu_e$ intersects with surface $S$. The energy distribution of these intersecting $\nu_e$ corresponds directly to our signal's energy spectrum for {a neutrino} detector. Additionally, the angle between the direction of $\nu_e$ and the line from the Sun to their intersection gives the desired $\theta_{\mr{Sun}}$ distribution.

\begin{figure}[!t]
\centering
\includegraphics[width=1.0\columnwidth]{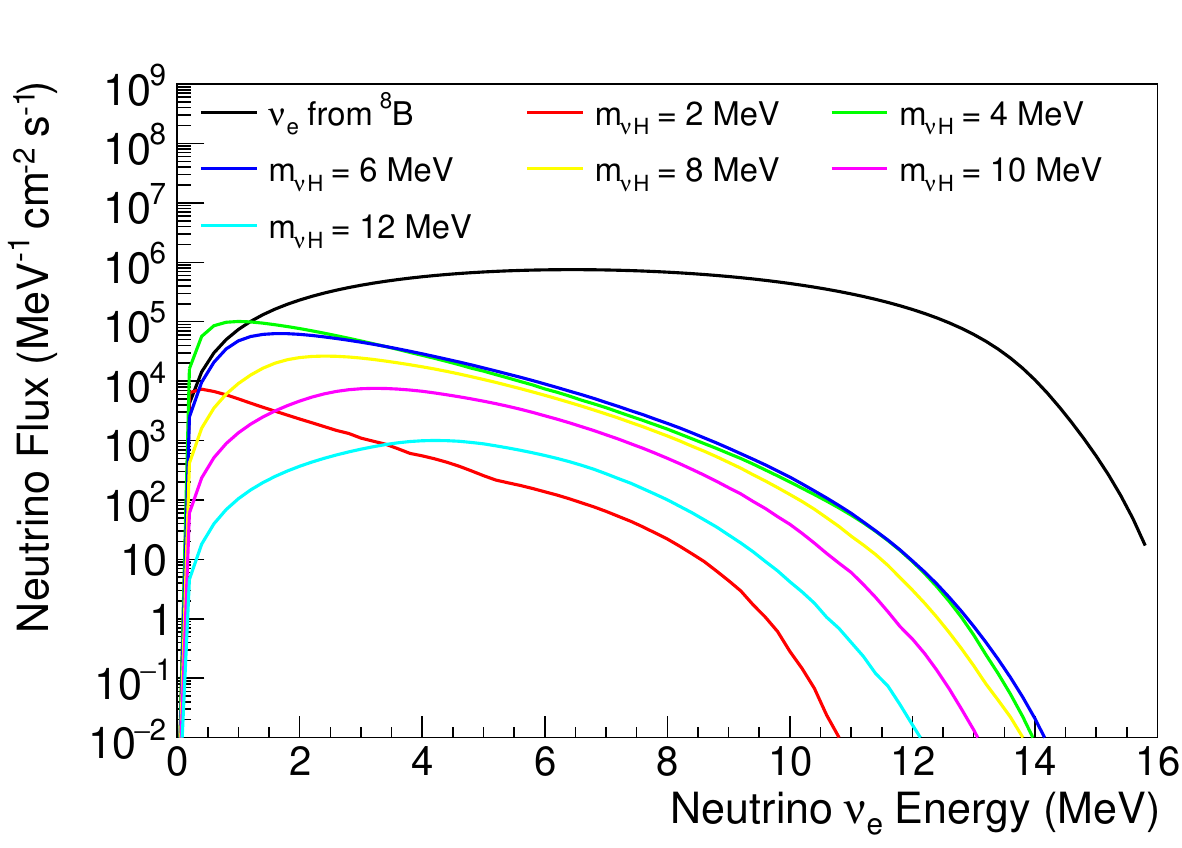}
\caption{{Energy spectra of $\nu_e$'s from $\nu_H$ decay that reach the detector on Earth for different values of $m_{\nu_H}$ with $|U_{eH}|^2 = 0.1$. The plot also shows the energy spectrum of background $^8 \mr{B}$ solar neutrino (black curve).}}
\label{fig:DecayInFlightNeutrinoEnergySpectrum}
\end{figure}

Fig.~\ref{fig:DecayInFlightNeutrinoEnergySpectrum} shows the energy spectra of $\nu_e$'s from $\nu_H$ decay that reach Earth’s detector. As expected, the signal $\nu_e$ from this decay is much softer than the background $^8 \mr{B}$ solar neutrino, peaking below $5~\mr{MeV}$. Distinguishing these low-energy $\nu_e$ from background solar neutrinos is challenging, as the electron energy from the elastic scattering of $^8 \mr{B}$ also peaks at low energies, as shown in Fig.\ref{fig:eeSignalRateAndSpectrum}. A good signal-to-background ratio will significantly enhance $\nu_e$ detection through charged-current interactions (as proposed in~\cite{liang2025LiCl,shao2023LiCl}), where the recoil energy closely correlates with the neutrino energy.

\begin{figure}[!t]
\centering
\includegraphics[width=1.0\columnwidth]{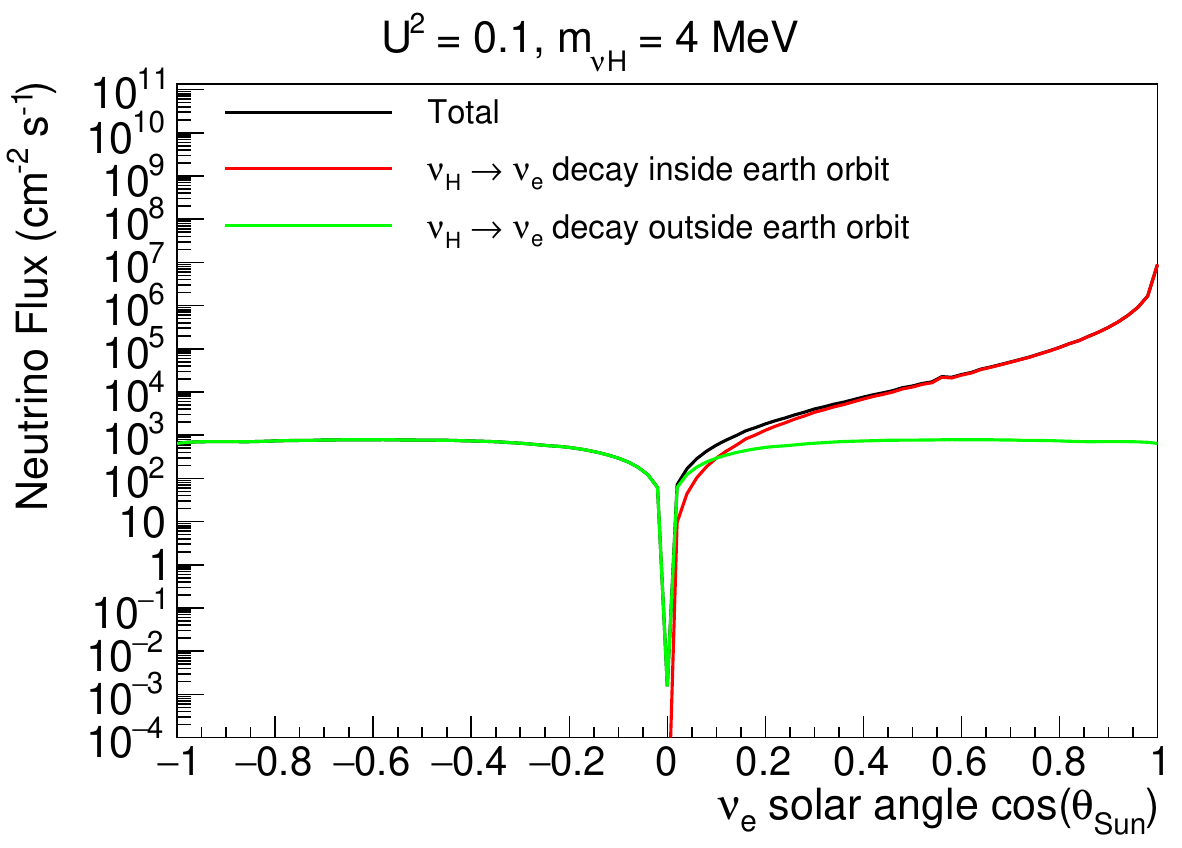}
\includegraphics[width=1.0\columnwidth]{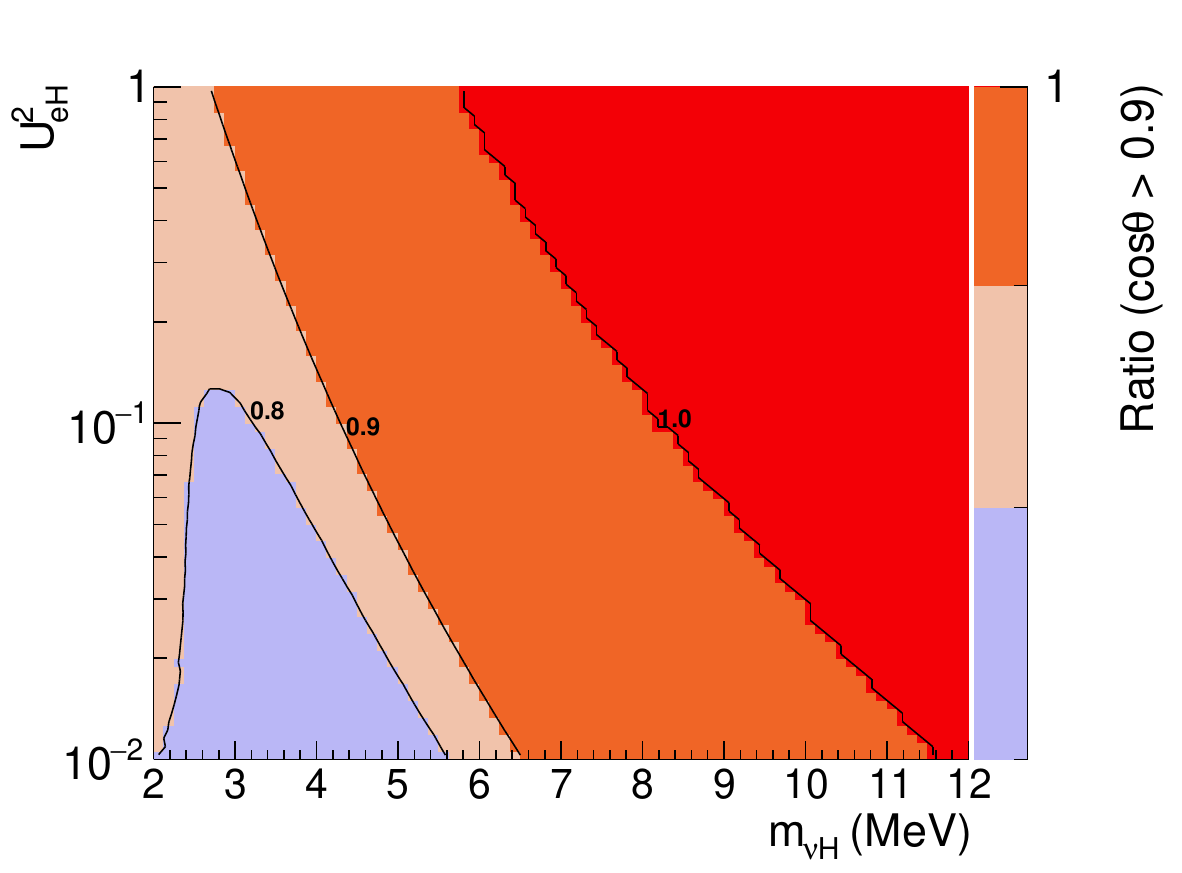}
\caption{Top: distribution of $\cos\theta_{\mr{Sun}}$ of $\nu_e$ from $\nu_H$ decay and then  reach the detector on earth with $m_{\nu_H} = \qty{4}{MeV}$ and $|U_{eH}|^2 = 0.1$. The distributions for $\nu_H$ decay inside and outside Earth orbit are shown separately. Bottom: Ratio of $\nu_e$ with $\cos\theta_{\mr{Sun}} > 0.9$ for different $m_{\nu_H}$ and $|U_{eH}|^2$.}
\label{fig:nueSolarAngle} 
\end{figure}

The solar angle $\theta_{\mr{Sun}}$ distinguishes the signal $\nu_e$ (from $\nu_H$ decay) from the solar neutrino background. The background solar neutrinos have a solar angle of zero since they originate from the Sun. Fig.\ref{fig:nueSolarAngle} illustrates the distribution of the solar angle for signal $\nu_e$ based on specific values of $m_{\nu_H}$ and $|U_{eH}|^2$. For signal $\nu_e$, there is a small but significant tail with large solar angles that can exceed typical angular resolution limits (around $25^{\circ}$) of standard detectors. The bottom plot in Fig.\ref{fig:nueSolarAngle} shows that for $|U_{eH}|^2>10^{-2}$, there is a large region with at least 10\% of $\nu_e$ events having a solar angle $\theta_{\mr{Sun}} > 25^{\circ}$ (or $\cos\theta_{\mr{Sun}} < 0.9$), allowing them to be distinguished from background neutrinos by major detectors capable of direction measurement.

\begin{figure}[!t]
\centerline{\includegraphics[width=1.0\columnwidth]{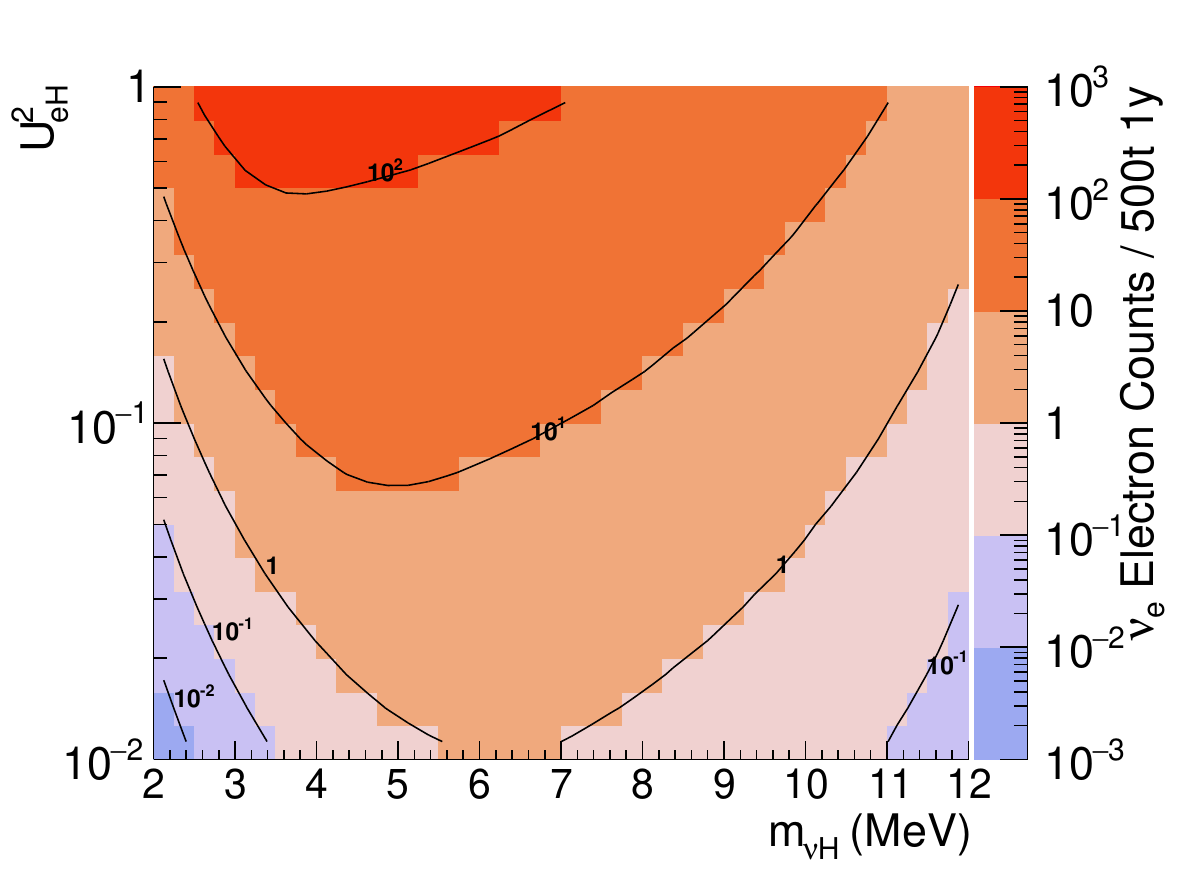}}
\caption{Event rate of $\nu_e$-electron elastic scattering per year with a 500-ton detector for $\nu_e$ signal from $\nu_H$ decay.}
\label{fig:nueSignalCounts}
\end{figure}

Considering the $\nu_e$-electron scattering cross section (approximately $9.4\times10^{-45}~\mr{cm}^2$ for $E_{\nu_e} = 10~\mr{MeV}$), we can estimate the expected event count rate for detecting $\nu_e$ signals from $\nu_H$ decay, as shown in Fig.\ref{fig:nueSignalCounts}. In most of the phase space where $|U_{eH}|^2 > 10^{-2}$, we expect to observe at least a few $\nu_H$ events by detecting $\nu_e$ from its decay, using about one year of data from a 500-ton solar neutrino detector. Unlike Fig.\ref{fig:eeSignalRateAndSpectrum}, much of this phase space cannot be explored by searching for $e^+e^-$ signals due to the short lifetime of $\nu_H$.

\begin{figure}[!htbp]
    \centering
    \includegraphics[width=1.0\columnwidth]{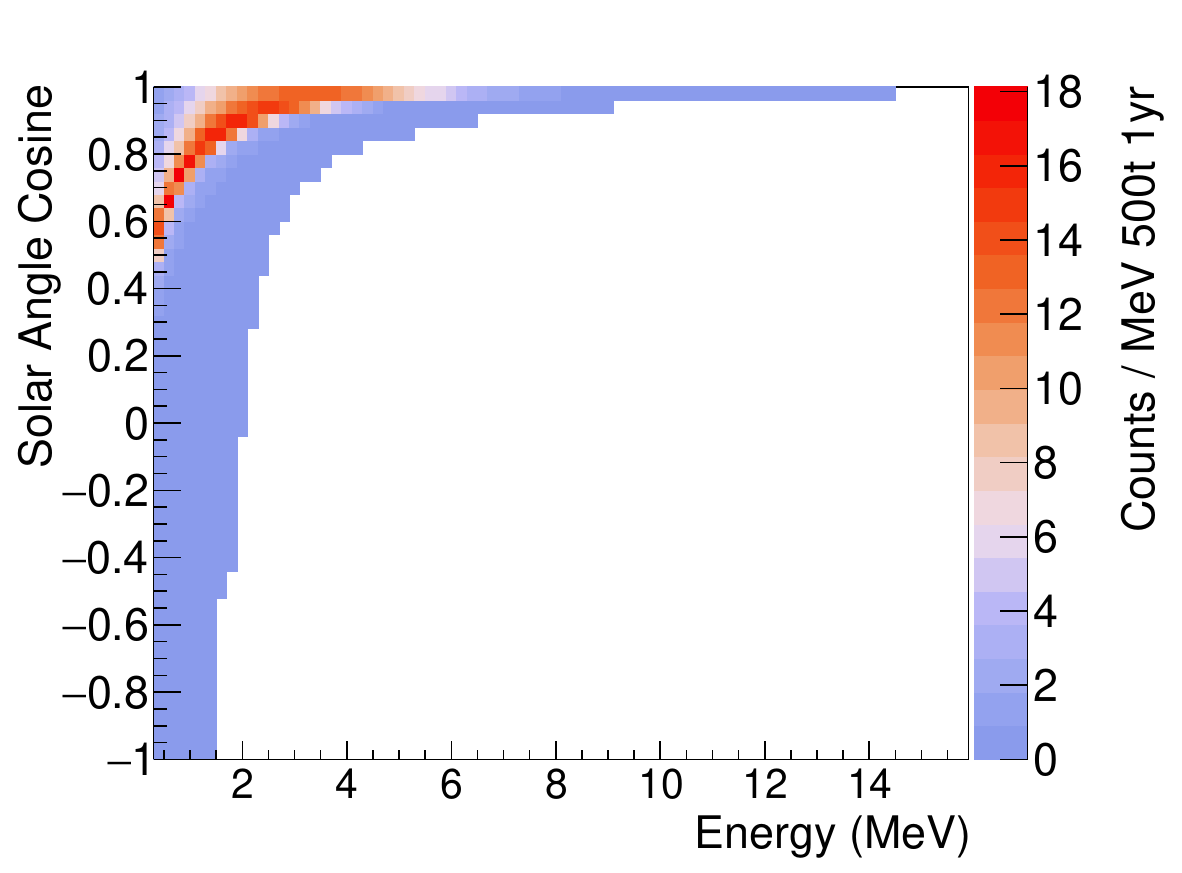}
    \includegraphics[width=1.0\columnwidth]{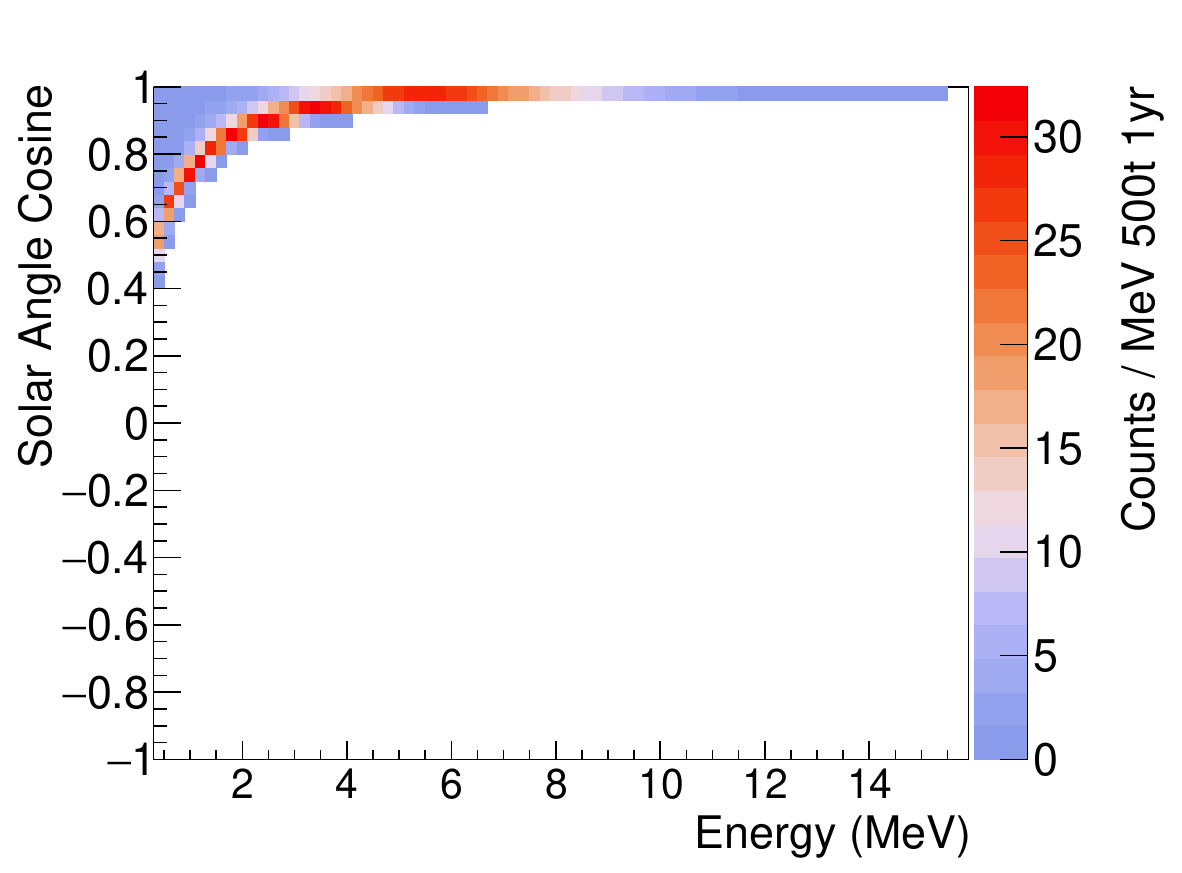}
    \caption{{The count rate distribution of solar angle cosine and scattering electron energy with $m_{\nu_H} = \qty{4}{MeV}$ and $|U_{eH}|^2=0.1$. Top: signal $\nu_e$ from $\nu_H$ decay; Bottom: $^8\mr{B}$ solar neutrino background. The background is confined to $\cos\theta_{\mr{Sun}} \gtrsim 0.4$ by the elastic-scattering kinematics, whereas the signal populates the full $\cos\theta_{\mr{Sun}}$ range, including the backward ($\cos\theta_{\mr{Sun}}<0$) region.}}
    \label{fig:s2_electron_2d_dist}
\end{figure}

In addition, the solar angle cosine distribution of $e^-$ from $\nu_e$-$e^-$ scattering is obtained and presented in Fig.\ref{fig:s2_electron_2d_dist}. 
Based on Fig.\ref{fig:s2_electron_2d_dist}, it is expected that scattered electrons with large solar angles are the low energy ones (below 2~$\mathrm{MeV}$ for $\cos(\theta_{\mr{Sun}})<0.9$).
At such low kinetic energies, it remains challenging to utilize directional information to effectively distinguish these signals from the $^8\mathrm{B}$ solar neutrino background.

\section{Sensitivities}
\label{sec:sensitivities}
To quantify the discovery potential across the parameter space, we evaluate the expected sensitivity using the profile likelihood method with asimov data sets~\cite{AsymptoticFormulaeLikelihoodbased2011}.

For method 1, the signal $S_i$ is the $e^+e^-$ energy spectrum from $\nu_H\to\nu_e e^+e^-$ decays inside the detector volume, while the background $B_i$ is the $^8\mr{B}$ solar neutrino--electron elastic scattering spectrum. Both the signal and background are convolved with a Gaussian energy resolution of $5\%/\sqrt{E[\mr{MeV}]}$ to account for detector response.

We construct an asimov data set $n_i = B_i$ and define the binned Poisson likelihood:

\begin{equation}
  \label{eq:likelihood}
\mathcal{L}(\mu, {\beta}) = \prod_i
\mathrm{Poisson}\big(n_i \big| \mu S_i + \beta B_i\big),
\end{equation}
where $\mu$ is the signal strength and $\beta$ is an unconstrained nuisance parameter that absorbs the systematic uncertainty in the solar neutrino flux. {The energy spectra are binned with a width of $0.2~\mr{MeV}$, and the fit is restricted to the window $4.8 < E_{ee} < 12.8~\mr{MeV}$ (40 bins) to suppress the environmental and cosmogenic backgrounds (e.g.\ $\ce{^{208}Tl}$ and $\ce{^{11}C}$) that dominate at lower energies.}

\begin{figure}[!htbp]
    \centering
    \includegraphics[width=1.0\columnwidth]{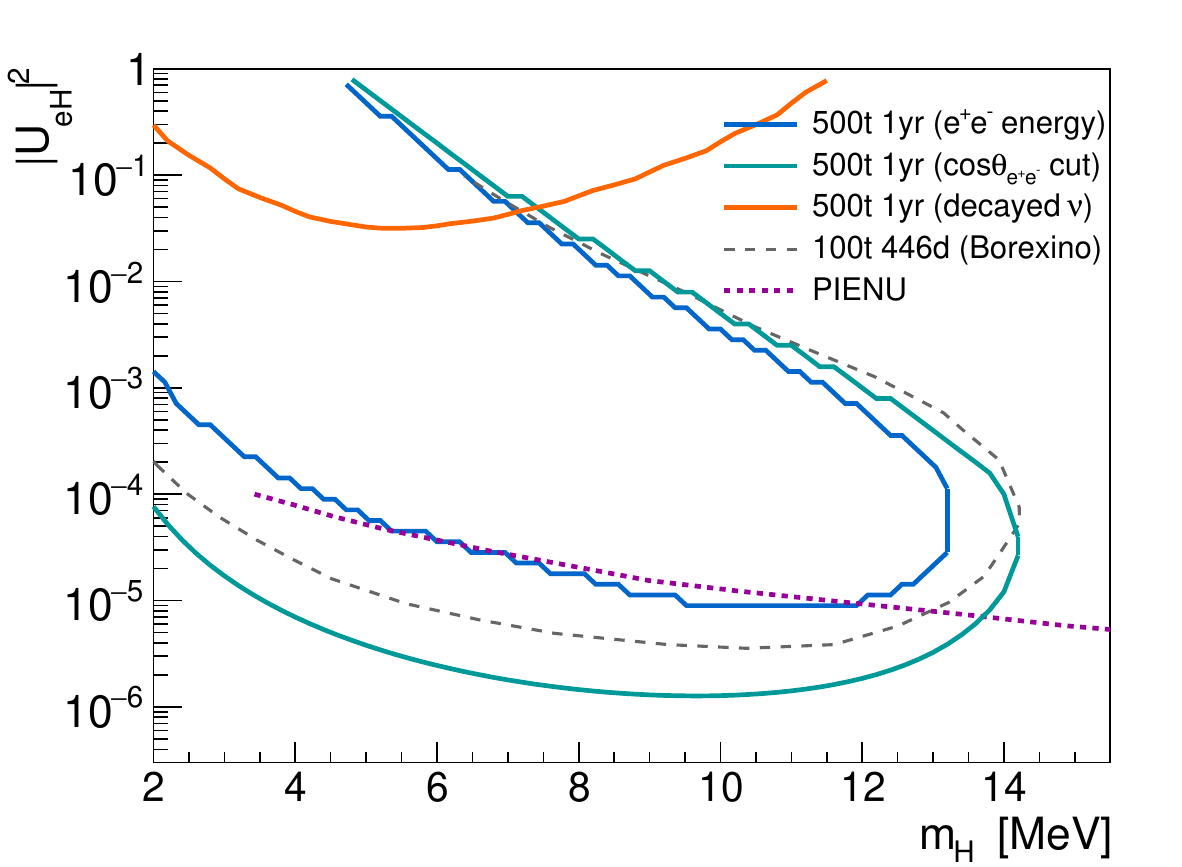}
    \caption{{Expected 90\% C.L.\ exclusion contours for a 500-ton detector with one year of exposure. The solid blue curve shows the sensitivity from method~1 using the $e^+e^-$ energy spectrum alone. The solid cyan curve shows the improved sensitivity from method~1 using the opening angle $\theta_{e^{+}e^{-}}$ with the $\cos\theta_{e^+e^-}<0.9$ selection. The solid orange curve shows the sensitivity from method~2, searching for $\nu_e$ from $\nu_H$ decay via elastic scattering. The gray dashed curve is the published Borexino exclusion~\cite{BorexinoRHN2013}. The magenta dotted curve shows the PIENU constraint~\cite{ImprovedSearchHeavy2018}.}}
    \label{fig:exclusion}
\end{figure}

For each point $(m_{\nu_H}, |U_{eH}|^2)$, we test the nominal signal hypothesis ($\mu=1$) against the background-only hypothesis ($\mu=0$) via the profile likelihood ratio~\cite{AsymptoticFormulaeLikelihoodbased2011}:

\begin{equation}
    \label{eq:likelihoodRatio}
q_1 = -2\ln \frac{\mathcal{L}({1}, \hat{\beta}_{{1}})}{\mathcal{L}({0}, \hat{\beta}_{{0}})},
\end{equation}
where $\hat{\beta}_\mu$ is the maximum likelihood estimate of $\beta$ for a fixed $\mu$. Under Wilks' theorem, $q_1$ follows a $\chi^2$ distribution with one degree of freedom. A parameter point is excluded at 90\% C.L.\ if $q_1 > 2.71$. The resulting expected exclusion contour is shown as the solid blue line in Fig.\ref{fig:exclusion}.

To further improve the sensitivity, we also examine the last analysis in section~\ref{sec:Method1}, which exploits the opening angle $\theta_{e^+e^-}$ between the electron and positron from $\nu_H$ decay. Neutrino detectors with directional sensitivity such as JNE can achieve an angular resolution of approximately $25^\circ$\footnote{{For JNE, a slow liquid-scintillator detector, the direction is reconstructed from the Cherenkov component of the light; in the simulations of Ref.~\cite{Luo_2023}, an angular resolution of better than $25^\circ$ is reached at electron energies above $6~\mr{MeV}$, the exact value depending on the scintillator. The resolution improves at higher energies. A fixed $25^\circ$ resolution is adopted for sensitivity estimates in this work.}}, corresponding to $\cos\theta_{e^+e^-} \lesssim 0.9$. Therefore, the $e^+e^-$ opening angle distribution is smeared with a Gaussian angular resolution of $25^\circ$ to account for the detector's directional response. A selection of $\cos\theta_{e^+e^-} < 0.9$ is also applied, and the corresponding improved exclusion sensitivity is shown as the solid cyan line in Fig.\ref{fig:exclusion}.

For method~2, which searches for $\nu_e$ from $\nu_H$ decay via $\nu_e$--electron elastic scattering, the same statistical approach is applied using a two-dimensional template fit in reconstructed electron energy $E_e$ and solar angle $\cos\theta_\mathrm{Sun}$. The signal includes $\nu_e$ from $\nu_H$ decays that undergo elastic scattering in the detector. The background is the standard $^8\mr{B}$ solar neutrino elastic scattering, with the component that oscillated into $\nu_H$ subtracted for consistency. The energy spectrum and angular distribution are both smeared in a similar way as method~1. {The two-dimensional template is binned in reconstructed electron solar angle $\cos\theta_{\mr{Sun}}$ (50 uniform bins over $[-1,1]$) and energy $E_e$ ($0.2~\mr{MeV}$ bins spanning the full range).} The expected 90\% C.L.\ exclusion contour from method~2 is shown as the solid orange line in Fig.\ref{fig:exclusion}.

For comparison, the Borexino constraint, which has already excluded a portion of the parameter space using 446~days of data~\cite{BorexinoRHN2013}, is also shown as the gray dashed curve in Fig.\ref{fig:exclusion}. {Complementary constraints in this mass range have also been reported from laboratory $\pi\to e\nu$ decay searches (PIENU)~\cite{ImprovedSearchHeavy2018,ImprovedConstraintsSterile2019}, which probe $\nu_H$ produced in pion decay rather than in the Sun.}

\section{Summary}
\label{sec:summary}

Solar neutrino experiments, particularly those with new designs that provide good energy and direction measurements, are promising for heavy sterile neutrino searches in the MeV mass range. 
In this range, $\nu_H$ can decay into an $e^+e^-$ pair plus an active neutrino $\nu_e$. 
This paper presents two methods based on this decay: one focuses on detecting ample $e^+e^-$ signals from $\nu_H$ decays with an intermediate lifetime, while the other aims to find $\nu_e$ from short-lived $\nu_H$ decays. 
The estimated signal and background event yields indicate the complementary sensitivity of both methods. 
Key variables to distinguish the $\nu_H$ signal from solar neutrino events are proposed along with their distributions. 
By combining these methods, it is expected to be sensitive in most of the phase space where $|U_{eH}|^2 > 10^{-6}$ and $2~\mr{MeV}<m_{\nu_H} < 15~\mr{MeV}$ using data from a 500-ton detector over one year.

\acknowledgments{}
This work is supported by the National Key Research and Development Plan of China (Project No. 2022YFA1604700), the National Natural Science Foundation of China (Grant No. 12521007 and No. 12575108), and the Tsinghua University Initiative Scientific Research Program.

\bibliography{main.bib}

\end{document}